\documentclass[conference, a4paper]{IEEEtran}
\usepackage{setspace}
\usepackage{clrscode}
\usepackage{amsmath}
\usepackage{amssymb}
\usepackage[dvips]{graphicx}
\usepackage{epsfig}
\usepackage{hyperref}
\usepackage{bm}
\usepackage{subfigure}
\usepackage{verbatim}
\usepackage{titletoc}
\usepackage{extarrows}
\usepackage{fancyhdr}
\usepackage{booktabs}
\usepackage{mathrsfs}

\usepackage{lipsum}
\usepackage{cuted}
\usepackage{stfloats}

\usepackage{cases}

\usepackage{algorithm}
\usepackage{algorithmic}

\usepackage{balance}

\newcommand{\teq}{\triangleq}
\newcommand{\mbf}{\mathbf}
\newcommand{\mbb}{\mathbb}
\newcommand{\mcal}{\mathcal}
\newcommand{\tbf}{\textbf}
\newcommand{\s}{\scriptstyle}
\newcommand{\sss}{\scriptscriptstyle}
\newcommand{\rH}{\rm{H}}
\newcommand{\rT}{\rm{T}}
\newcommand{\mH}{\mathbf{H}}
\newcommand{\tr}{\text{tr}}
\newcommand{\yk}{\mbf{y}_k}
\newcommand{\mG}{\mbf{G}}
\newcommand{\hG}{\widehat{\mbf{G}}}
\newcommand{\hH}{\widehat{\mbf{H}}}
\newcommand{\hHk}{\widehat{\mbf{H}}_k}
\newcommand{\Hdk}{\mbf{H}_{d,k}}
\newcommand{\edk}{\sigma^2_{d,k}}
\newcommand{\hHdk}{\widehat{\mbf{H}}_{d,k}}
\newcommand{\Hrk}{\mbf{H}_{r,k}}
\newcommand{\erk}{\sigma^2_{r,k}}
\newcommand{\eg}{\sigma^2_{g}}
\newcommand{\hHrk}{\widehat{\mbf{H}}_{r,k}}

\newcommand{\sedk}{\triangle \mbf{H}_{d,k}}
\newcommand{\serk}{\triangle\mbf{H}_{r,k}}
\newcommand{\seG}{\triangle\mbf{G}}

\newcommand{\edi}{\sigma^2_{d,i}}

\newcommand{\eri}{\sigma^2_{r,i}}
\newcommand{\hHri}{\widehat{\mbf{H}}_{r,i}}
\newcommand{\mW}{\mathbf{W}}
\newcommand{\tW}{\widetilde{\mW}}
\newcommand{\bW}{\mbf{\overline{W}}}
\newcommand{\mlW}{\bW}
\newcommand{\IRS}{\bm{\Theta}}

\newcommand{\Ek}{\mbf{E}_k^{*}}
\newcommand{\Tk}{\mbf{T}_k^{*}}
\newcommand{\Ti}{\mbf{T}_i^{*}}
\newcommand{\Ck}{\mbf{C}_k^{*}}
\newcommand{\Ci}{\mbf{C}_i^{*}}

\newcommand{\nCopt}{\mbf{C}_k}
\newcommand{\nTopt}{\mbf{T}_k}
\newcommand{\nWopt}{\bW}
\newcommand{\ntWopt}{\tW}

\newenvironment{sequation}{\begin{equation}\small}{\end{equation}}

\newtheorem{Lem1}{\tbf{Proposition}}

\begin{document}

\title{Joint Beamforming Design for IRS-Aided Communications with Channel Estimation Errors}
    \author{
\IEEEauthorblockN{Piao Zeng\IEEEauthorrefmark{1}, Deli Qiao\IEEEauthorrefmark{1}\IEEEauthorrefmark{2}, and Haifeng Qian\IEEEauthorrefmark{3}}
\IEEEauthorblockA{\IEEEauthorrefmark{1}\small{School of Communication and Electronic Engineering, East China Normal University, Shanghai, China}}
\IEEEauthorblockA{\IEEEauthorrefmark{2}\small{National Mobile Communications Research Laboratory, Southeast University, Nanjing, China}}
\IEEEauthorblockA{\IEEEauthorrefmark{3}\small{School of Software Engineering, East China Normal University, Shanghai, China}}
\small{Email: 52181214005@stu.ecnu.edu.cn, dlqiao@ce.ecnu.edu.cn, hfqian@cs.ecnu.edu.cn}}

\maketitle

\begin{abstract}\let\thefootnote\relax\footnotetext{This work is supported in part by the National Natural Science Foundation of China (61671205), and in part by the open research fund of National Mobile Communications Research Laboratory, Southeast University (2020D02).}
This paper investigates the joint design of the beamforming scheme in intelligent reflecting surface (IRS) assisted multiuser (MU) multiple-input multiple-output (MIMO) downlink transmissions. Channel estimation errors associated with the minimum mean square error (MMSE) estimation are assumed and the weighted sum rate (WSR) is adopted as the performance metric. Low-resolution phase shifters (PSs) in practical implementations are taken into account as well. Under the constraint of the transmit power and discrete phase shifters (PSs), an optimization problem is formulated to maximize the WSR of all users. To obtain the optimal beamforming matrices at the IRS, two solutions based on the majorization-minimization (MM) and successive convex approximation (SCA) methods, respectively, are proposed. Through simulation results, both of the proposed two schemes achieve a significant improvement in WSR. Furthermore, the superiority of the SCA-based solution is demonstrated. Overall, two viable solutions to the joint beamforming design in IRS-aided MU-MIMO downlink communication systems with channel estimation errors are provided.
\end{abstract}

\section{Introduction}

Thanks to the recent advances in metamaterials and microelectro-mechanical systems (MEMS), intelligent reflecting surface (IRS) has stood out as an effective complementary medium to support the existing wireless communication systems \cite{Intro_MEMS}. With a thin planar composed of massive reconfigurable passive elements, IRS can modify the phase shifts (PS) of the incident signals in a software-controlled fashion, boosting the received signal power and suppressing the interference as well with extremely low power consumption, which improves the communication capacities \cite{Intro_qqw}. Benefiting from its flexibility in deployment, the IRS is expected to be wildly put into use, which arouses the extensive and in-depth discussion and research in both the industry and the academia \cite{WSR_hyg}--\cite{CR}.

Although numerous efforts have been invested in this area, most prior works focused on the multiple-input single-output (MISO) systems \cite{Intro_qqw}--\cite{qqw_dis}. For instance, in \cite{SWIPT}, the authors studied the rate-energy performance trade-off of the simultaneous wireless information and power transfer (SWIPT) system, where two sets of single-antenna receivers were considered. In \cite{CR}, the authors presented the robust beamforming design for the IRS-aided cognitive radio (CR) systems with single-antenna primary users and secondary users (SUs). Generally, the optimization problem in such MISO systems can be formulated as a quadratically constrained quadratic program (QCQP) and solved with the change of variables for the phases of PSs \cite{Intro_qqw}, which are inapplicable to multi-antenna users, i.e., the MIMO scenario. Specifically, the increased antenna arrays deployed at both transceivers and IRS expand the dimensionality of the associated channel matrices, which hinder the aforementioned transformation of variables.

Another noteworthy issue is that, the well-designed IRS-assisted system is based on the accurate channel state information (CSI). Despite various studies have investigated the channel estimation in the IRS-aided communication systems and provided several effective approach \cite{Intro_CE1}--\cite{Intro_CE3}, the estimation errors are still inevitable in most cases. Hence, it is meaningful to take these losses into account to explore the full potential of the IRS. Nonetheless, only few works are related to the beamforming design with channel estimation errors, which are basically designed for single-antenna users \cite{WSR_hyg}, \cite{CR}, \cite{imCSI1}, \cite{imCSI2}.

In this paper, we consider the beamforming design in IRS assisted multiuser (MU) MIMO downlink transmissions. We assume that there are channel estimation errors for each link, and users are equipped with multiple antennas. We first formulate an optimization problem that maximizing the weighted sum rate (WSR) of all the users. Next, we transform the original WSR maximization problem into an equivalent weighted minimum mean square error (WMMSE) minimization problem and decompose it into two sub-problems, namely active and passive beamforming. Subsequently, we solve these two sub-problems alternatively. In particular, we tackle the active beamforming problem with the Lagrange multipliers method. While for passive beamforming, we propose two solutions by utilizing the majorization-minimization (MM) and successive convex approximation (SCA) techniques, respectively. Through numerical evaluations, we validate the superiority and effectiveness of the proposed algorithms.

\begin{figure*}[ht]
\vspace*{1pt}
\begin{sequation}
\begin{aligned}
\yk=&\big[\hHdk^{\rH}+\sedk^{\rH}+(\hHrk^{\rH}+\serk^{\rH})\IRS(\hG+\seG)\big]\sum\nolimits_{k=1}^{K} \mbf{W}_{k} \mbf{s}_{k}+\mbf{n}_k \\
=&\hHk\mbf{W}_{k} \mbf{s}_{k}+
\underbrace{
\hHk\sum\nolimits_{i\neq k}^K \mbf{W}_{i} \mbf{s}_{i}+(\sedk^{\rH}+\hHrk^{\rH}\IRS\seG+\serk^{\rH}\IRS\hG+\serk^{\rH}\IRS\seG)\sum\nolimits_{i=1}^{K} \mbf{W}_{i} \mbf{s}_{i}+\mbf{n}_k .
}_{\text{(a) Interference and noise}}
\end{aligned}
\label{eq:yk2}
\tag{2}
\end{sequation}
\hrulefill
\end{figure*}

The paper is organized as follows. Section II briefly discusses the system model. Section III presents the main contributions of this work including the formulation and solution of the optimization problem. The simulation results are provided in Section IV. Finally, Section V concludes the paper.

\emph{Notations:} Throughout the paper, superscripts $(\cdot)^{\mathrm{T}}$, $(\cdot)^{\mathrm{H}}$, and $(\cdot)^{-1}$ represent the transpose, Hermitian transpose, and inverse of a matrix, respectively. tr$(\mbf{X})$, $|\mbf{X}|$ and $\|\mbf{X}\|$ denote the trace, determinant and Frobenius norm of matrix $\mbf{X}$, respectively. $\mbb{C}^{a \times b}$ expresses the space of $a \times b$ complex matrices. $\mathbf{I}_{n}$ (sometimes the subscript $n$ is omitted) stands for the $n \times n$ identity matrix. $\mathcal{C}\mathcal{N}\left(\bm{\mu}, \mbf{R}\right)$ denotes circularly symmetric complex gaussian (CSCG) random distributions with mean $\bm{\mu}$ and covariance matrix $\mbf{R}$. $\nabla f$ denotes the gradient of the function $f$. $\text{Re}\{\cdot\}$ represents the real part of a complex value. $\mathbb{E}[\cdot]$, diag$(\cdot)$, $\otimes$, and $\circ$ are the expectation operator, diagonalization operator, Kronecker product operator and Hadamard product operator, respectively.

\section{System Model}

In the subsequent analysis, we consider an IRS-aided downlink MU-MIMO communication system, which consists of one BS, one IRS and $K$ users. We assume that the BS is equipped with $M$ transmit antennas, the IRS has $N$ reflecting elements and each user is equipped with $N_r$ receive antennas. We assume that the reflections of the signals via the IRS for more than once are omitted and adopt the quasi-static flat-fading model for the channels \cite{Intro_qqw}. Denote the set of users and reflecting elements as $\mathcal{K}\teq\{1,\cdots,K\}$ and  $\mathcal{N}\teq\{1,\cdots,N\}$, respectively. The analysis in the sequel is for $\forall k \in \mcal{K}$ and $\forall n \in \mcal{N}$, if not specified otherwise.

The received signal at the $k^{th}$ user is given by
\begin{sequation}
\setcounter{equation}{1}
\begin{aligned}
\yk&=(\Hdk^{\rH}+\Hrk^{\rH}\IRS\mG) \sum\nolimits_{i=1}^{K} \mbf{W}_{i} \mbf{s}_{i}+\mbf{n}_k .
\end{aligned}
\label{eq:yk}
\end{sequation}
In this equation, $\mathbf{H}_{d,k} \in\mathbb{C}^{M \times N_r}$, $\mathbf{H}_{r,k} \in\mathbb{C}^{N \times N_r}$, $\mathbf{G} \in\mathbb{C}^{N \times M}$ represent the channel matrix from the BS to the $k^{th}$ user, from the IRS to the $k^{th}$ users and from the BS to the IRS, respectively. $\mbf{\Theta}=\text{diag}(\eta_1e^{j\phi_1},\cdots,\eta_n e^{j\phi_n},\cdots,\eta_Ne^{j\phi_N})$ is defined as the diagonal reflection coefficients matrix adopted at the IRS,  where $\eta_n \in [0,1]$ and $\phi_n \in [0,2\pi)$ are the reflection coefficient amplitude and PS of the $n ^{th}$ reflecting element, respectively. In this paper, we set $\eta_n=1,\forall n \in \mathcal{N}$ to maximize the signal reflection \cite{qqw_dis}, and focus on the design of the PS $\phi_n$ of each element in the matrix $\IRS$. Note that, considering the practical implementation constraints, the value of each PS can only be chosen from a finite set, which is defined as $\mcal{F}\teq\{0,\frac{2\pi}{L},\cdots,\frac{2\pi(L-1)}{L}\}$ with $L=2^B$, where $B$ is the number of bits for the quantization \cite{qqw_dis}. $\mathbf{s}_k \in\mathbb{C}^{N_s \times 1} \sim \mathcal{C}\mathcal{N}\left(\mbf{0}, \mbf{I}_{N_s}\right)$ represents the $N_s$ independent and identically distributed (i.i.d.) desired data streams for the $k^{th}$  user, which  are precoded by
the beamforming matrix $\mbf{W}_{k}\in \mbb{C}^{M \times N_{s}}$. $\mbf{n}_k\sim\mathcal{C}\mathcal{N}\left(\bm{0}, \sigma_{k}^{2}\mbf{I}_{N_r}\right)$ is the received addictive white Gaussian noise (AWGN) at the $k^{th}$ user with $\sigma_{k}^{2}$ denoting the noise power at each antenna of the $k^{th}$ user.

For practical considerations, obtaining the accurate CSI is challenging, especially for the channel estimation from the IRS to users due to the mobility of the users and passive property of the IRS. Hence, the channel estimation errors is inevitable. Specifically, the actual channel is composed of the estimated CSI and corresponding CSI errors, i.e.,  $\mbf{G}=\mbf{\widehat{G}}+\seG,\mbf{H}_{d,k} =\mbf{\widehat{H}}_{d,k}+\sedk, \mbf{H}_{r,k}=\mbf{\widehat{H}}_{r,k}+\serk$.
According to \cite{hyg_CE43}, \cite{hyg_CE44}, the channel estimation errors are assumed to be uncorrelated with the estimated channel coefficients by the minimum mean square error (MMSE) estimation. Thus, we model the true channel $\mH$ as $\mH \sim \mathcal{C}\mathcal{N}\big(\widehat{\mH}, \mbf{A} \otimes \mbf{B}\big)$, where the estimated channel matrix $\widehat{\mH}$ is its mean value, $\mbf{A}$ and $\mbf{B}$ are the covariance matrix seen from the receiver side and the transmitter side, respectively. In particular, $\Hdk \sim \mathcal{C}\mathcal{N}\big(\hHdk, \mbf{A}_{d,k} \otimes \mbf{B}_{d,k}\big)$, $\Hrk \sim \mathcal{C}\mathcal{N}\big(\hHrk, \mbf{A}_{r,k} \otimes \mbf{B}_{r,k}\big)$ and $\mG \sim \mathcal{C}\mathcal{N}\big(\hG, \mbf{A}_{g} \otimes \mbf{B}_{g}\big)$. And we assume that, $\mbf{A}_{d,k}=a_{d,k}\cdot\mbf{I}$, $\mbf{B}_{d,k}=b_{d,k}\cdot\mbf{I}$, $a_{d,k}\cdot b_{d,k}=\sigma^2_{d,k}$, $\mbf{A}_{r,k}=a_{r,k}\cdot\mbf{I}$, $\mbf{B}_{r,k}=b_{r,k}\cdot\mbf{I}$, $a_{r,k}\cdot b_{r,k}=\sigma^2_{r,k}$, $\mbf{A}_{g}=a_{g}\cdot\mbf{I}$, $\mbf{B}_{g}=b_{g}\cdot\mbf{I}$ and $a_{g}\cdot b_{g}=\sigma^2_{g}$.

Under this model, we recast the received signal at the $k^{th}$ user by ($\ref{eq:yk2}$) at the top of this page, where $\hHk=\hHdk^{\rH}+\hHrk^{\rH}\IRS\hG$.
For notation simplicity, denote (a) in (\ref{eq:yk2}) as $\mbf{N}_k\in \mbb{C}^{N_r \times 1}$, $\bW\teq[\mbf{W}_{1},\mbf{W}_{2},\cdots,\mbf{W}_{K}]\in \mbb{C}^{M \times N_s K}$, and $\tW \triangleq \sum_{i=1}^{K} \mbf{W}_{i}\mW_{i}^{\rH} \in \mbb{C}^{M \times M}$.

% Proposition 1
\begin{Lem1}\label{prop:rate}
The achievable WSR is given by \cite{WMMSE}
\begin{sequation}
\setcounter{equation}{3}
\begin{aligned}
     \mcal{R} &= \sum\nolimits_{k=1}^K \omega_k\text{log}_2 \left|\mbf{I}+ \hHk \mW_k \mW_k^{\rH} \hHk^{\rH}  \mbf{J}_k^{-1}  \right|,
\end{aligned}
\end{sequation}
where $\mbf{J}_{k} \triangleq \mbb{E}\left\{ \mbf{N}_k \mbf{N}_k^{\rH} \right\}$ is defined as the covariance matrix of the interference and noise at the $k^{th}$ users:
\begin{sequation}
\begin{aligned}
\mbf{J}_{k}
= &\sum_{i\neq k}^K \hHk \mW_{i} \mW_{i}^{\rH} \hHk^{\rH}
+\eg\tr(\tW) \cdot \hHrk^{\rH}\hHrk
+\alpha_{k} \mbf{I},
\end{aligned}
\label{eq:Jk}
\end{sequation}
where $\alpha_{k} = (\edk+N\eg\erk)\cdot\tr(\tW)+\erk\tr( \hG\tW \hG^{\rH} )+\sigma_{k}^2$.
\end{Lem1}

\emph{Proof:} Expanding $\mbf{J}_{k}$ with (\ref{eq:yk2}), we have
\begin{sequation}
\begin{aligned}
\mbf{J}_{k}
= &
\mbb{E}\big\{ \sum\nolimits_{i\neq k}^K \hHk \mW_{i} \mW_{i}^{\rH} \hHk^{\rH} \big\}
+\mbb{E}\big\{ \sedk^{\rH} \tW \sedk \big\} \\
+& \mbb{E}\big\{ \hHrk^{\rH}\IRS\seG \tW (\hHrk^{\rH}\IRS\seG)^{\rH} \big\} \\
+&\mbb{E}\big\{ \serk^{\rH}\IRS\hG \tW  (\serk^{\rH}\IRS\hG)^{\rH} \big\} \\
+& \mbb{E}\big\{ \serk^{\rH}\IRS\seG \tW (\serk^{\rH}\IRS\seG)^{\rH} \big\}
+\sigma_{k}^2 \mbf{I}.
\end{aligned}
\label{eq:Jk1}
\end{sequation}
Using the lemma in \cite{ImCSI}: for $\mH \sim \mathcal{C}\mathcal{N}\big(\widehat{\mH}, \mbf{A} \otimes \mbf{B}\big)$, there is $\mbb{E}_{\rm H}[{\bf H} {\bf X} {\bf H}^{\rH}]={\hH} {\bf X} {\hH}^{\rH}+{\rm tr}({\bf X}{\mbf{A}}^{\rT})\cdot{\mbf{B}}$, the second term in (\ref{eq:Jk1}) can be written as
\begin{sequation}
\begin{aligned}
\mbb{E}\big\{ \sedk^{\rH} \tW \sedk \big\}= \tr(\tW\mbf{A}_{d,k}^{\rT})\mbf{B}_{d,k}=\edk\tr(\tW)\cdot \mbf{I}.
\end{aligned}
\end{sequation}
Computing the rest terms in the same manner with this lemma, we can obtain $\mbf{J}_{k}$ as in (\ref{eq:Jk}), which completes the proof. \hfill$\square$

\section{Problem Formulation And Solution}
\subsection{Problem Formulation}
In this paper, our objective is to maximize the WSR of all the users by jointly designing the transmit beamforming at the BS and the PS matrix at the IRS, subject to the transmit power constraint and PS constraint, which can be expressed as
\begin{sequation}
{\mcal{P}(\text{A})} \hspace {0.3cm} {\underset{\scriptscriptstyle\mathbf{\bW,\Theta}}{\text{max}}} \hspace {0.2cm}  {\mcal{R}} \hspace {0.4cm}
{\text{s.t.}} \hspace {0.2cm}
\left\{
\begin{aligned}
&\hspace {0.1cm}{\sum\nolimits_{k=1}^K \|\mbf{W}_{k}\|^2\leq P_t, }  \\
&\hspace {0.1cm}{\phi_n \in \mcal{F}}, \hspace {0.2cm} \forall n \in \mcal{N}.
\end{aligned}
\right.
\end{sequation}

To tackle the WSR maximization problem, one popular and effective approach is to transform the original problem into a WMMSE minimization problem \cite{WMMSE}. To fulfill the transformation, firstly we introduce a hypothetical receive filter $\mbf{C}_k$ at the $k^{th}$ user, and calculate the MSE matrix $\mbf{E}_k$ as
\begin{sequation}
\begin{aligned}
     \mbf{E}_k =& \mbb{E}\left[ (\mbf{C}_k^{\rH} \mbf{y}_k-\mbf{s}_k)(\mbf{C}_k^{\rH} \mbf{y}_k-\mbf{s}_k)^{\rH} \right] \\
     =&\mbf{C}_k^{\rH} \mbf{Q}_k \mbf{C}_k
     -\mbf{C}_k^{\rH} \hHk \mbf{W}_{k}
     -\mbf{W}_k^{\rH} \hHk^{\rH}  \mbf{C}_{k}
     +\sigma_{k}^2 \mbf{C}_k^{\rH} \mbf{C}_k+\mbf{I},
\label{eq:MSE}
\end{aligned}
\end{sequation}
where $\mbf{Q}_k$ is given by
\begin{sequation}
\begin{aligned}
\mbf{Q}_{k}
\teq& \hHk\tW\hHk^{\rH} +\edk\tr(\tW)\cdot \mbf{I}
+\eg\tr(\tW) \cdot \hHrk^{\rH}\hHrk \\
&+\erk\tr( \hG\tW \hG^{\rH} )\cdot \mbf{I}
+\eg\erk\cdot N \cdot \tr( \tW ) \cdot \mbf{I} .
\end{aligned}
\label{eq:MSE1}
\end{sequation}
By introducing another variables $\mbf{T}_k\in \mbb{C}^{N_s \times N_s}\succeq \bf{0}$, the original WSR maximization problem is equivalent to the following WMMSE minimization problem \cite{WMMSE}:
\begin{small}
\begin{align}
{\mcal{P}(\text{B})}  \min_{\underset{\{\mbf{C}_k\},\{\mbf{T}_k\}}{\sss\bW,\mathbf{\Theta},}} \hspace{0.1cm} & {f \triangleq \sum_{k=1}^K \left\{ \tr(\mbf{T}_k \mbf{E}_k)-\omega_k \log_2 | 1 / \omega_k   \cdot\mbf{T}_k|\right\}} \label{pro:B}\\
{\text{s.t.}} \hspace{0.6cm} & {\sum\nolimits_{k=1}^K \|\mbf{W}_{k}\|^2\leq P_t, } \tag{\ref{pro:B}{a}} \label{pro:Ba}\\
&\mbf{T}_k\succeq {\bf0}, \quad \forall k \in \mcal{K}, \tag{\ref{pro:B}{b}} \label{pro:Bb}\\
&{\phi_n \in \mcal{F}}, \quad n \in \mcal{N}. \tag{\ref{pro:B}{c}} \label{pro:Bc}
\end{align}
\end{small}

To handle the coupled variables in the optimization problem, we utilize the classical block coordinate descent (BCD) method \cite{BCD}. Specifically, to begin with, we decompose the original problem into two sub-problems, i.e., the active and passive beamforming optimization problems:
\begin{small}
\begin{align}
&\mcal{P}(\text{B1})\hspace{0.1cm} \underset{\sss\bW,\{\mbf{C}_k\},\{\mbf{T}_k\}}{\text{minimize}} \hspace{0.1cm}  {f\left(\bW,\{\mbf{C}_k\},\{\mbf{T}_k\}\right)} \hspace{0.1cm} \text{s.t.} \hspace{0.05cm}(\ref{pro:Ba}),(\ref{pro:Bb}),
\label{pro:B1} \\
&\mcal{P}(\text{B2})\quad \underset{\scriptscriptstyle \IRS}{\text{min}} \hspace{0.1cm}  {f\left(\IRS\right)} \hspace{0.3cm} \text{s.t.} \hspace{0.3cm} {\phi_n \in \mcal{F}}, \hspace{0.15cm} \forall n \in \mcal{N},
\label{pro:B2}
\end{align}
\end{small}
respectively.

Afterwards, these two sub-problems are solved separately and alternatively. In particular, first we optimize the transmit (active) beamforming problem with a fixed $\bm{\Theta}$. Then, we find the optimal $\bm{\Theta}$ with the obtained $\mlW^{*}$ and $\{\mbf{C}_k^{*}\},\{\mbf{T}_k^{*}\}$. This procedure will repeat until satisfying the stopping criteria. Note that, this alternating optimization approach has been widely used in the IRS-assisted system \cite{Intro_qqw}--\cite{SWIPT}, whose convergence and optimality have been discussed in \cite{imCSI1}, \cite{WMMSE}.

\subsection{Transmit Beamforming Optimization with Fixed PS Matrix}
In this section, we solve the first sub-problem with the giving $\IRS$. Still, the BCD method is adopted, where the variables $\{\mbf{T}_k\}$, $\{\mbf{C}_k\}$ and $\bW$ are updated alternatively with the others fixed.

The optimal $\mbf{C}_k$ is obtained by letting the first-order derivative of $\mbf{E}_k$ with respective to (w.r.t.) $\mbf{C}_k$ be zero:
\begin{sequation}
\begin{aligned}
\Ck &= \arg \underset{\mbf{C}_k}{\min} \hspace {0.2cm} \mbf{E}_k
= \big(\mbf{Q}_k+\sigma_{k}^2 \mbf{I} \big)^{-1} \hHk \mW_k.
\end{aligned}
\label{eq:com}
\end{sequation}

Substitute (\ref{eq:com}) in (\ref{eq:MSE}), the corresponding MSE matrix $\mbf{E}_k$ and $\mbf{T}_k$ are given by
\begin{sequation}
\begin{aligned}
\Ek &= \mbf{I} - \mW_k^{\rH} \hHk^{\rH} \Ck, \hspace{0.2cm}
\Tk=\omega_k (\Ek)^{-1}.
\end{aligned}
\label{eq:Ek_Tk}
\end{sequation}

As for the precoding matrix $\bW$, to cope with the transmit power constraint, we introduce a dual variable $\lambda\geq 0$ and find the optimal solution via Lagrange multipliers method. Given the obtained $\{\Ck\}$ and $\{\Tk\}$, the Lagrangian function of (\ref{pro:B1}) is constructed and simplified as
\begin{sequation}
\begin{aligned}
\mcal{L}(\bW,\lambda)&=\lambda \big[ \tr(\tW)-P_t\big]+\sum_{k=1}^K\tr \big\{ \Tk (\Ck)^{\rH} \mbf{Q}_k \Ck \\
-&\Tk (\Ck)^{\rH} \hHk \mbf{W}_{k}-\Tk\mbf{W}_k^{\rH} \hHk^{\rH}  \Ck
\big\} .
\end{aligned}
\label{eq:MSE1}
\end{sequation}

Letting the first-order derivative of $\mcal{L}(\bW,\lambda)$ w.r.t. $\mW_k$ be zero, we can obtain the optimal $\mW_{k}^*(\lambda)$ by \footnote{For brevity, the superscription of $\Ci$, $\Ck$, $\Ti$ and $\Tk$ are omitted in this equation, i.e., $\mbf{C}_i$, $\mbf{C}_k$, $\mbf{T}_i$ and $\mbf{T}_k$ actually refer to $\Ci$, $\Ck$, $\Ti$ and $\Tk$.}
\begin{sequation}
\begin{aligned}
&\mW_{k}^*(\lambda)=\bigg[\sum_{i=1}^{K}  \bigg({\hat{\mbf{H}}_i^{\rH} \mbf{C}_i \mbf{T}_i \mbf{C}_i^{\rH} \hat{\mbf{H}}_i}
+ \edi \cdot\tr(\mbf{T}_i \mbf{C}_i^{\rH} \mbf{C}_i )\mathbf{I} \\
&\hspace{0.2cm}+\eg \cdot\tr(\mbf{T}_i \mbf{C}_i^{\rH} \hHri^{\rH} \hHri\mbf{C}_i )\mathbf{I}
+\eri \cdot\tr(\mbf{T}_i \mbf{C}_i^{\rH} \mbf{C}_i ) \hG^{\rH} \hG \\
&\hspace{0.2cm}+\eg \eri \cdot N \cdot\tr(\mbf{T}_i \mbf{C}_i^{\rH} \mbf{C}_i ) \mathbf{I}
\bigg) +\lambda\mathbf{I}\bigg]^{-1}\hHk^{\rH} \mbf{C}_k \mbf{T}_k,
\end{aligned}
\label{eq:Wk}
\end{sequation}
where the optimal $\lambda$ is obtained by solving the dual problem $\underset{\scriptscriptstyle \lambda \geq 0}{\text{max}} \hspace{0.23cm} \underset{\scriptscriptstyle \bW}{\text{min}} \hspace{0.2cm} \mcal{L}(\bW,\lambda)$, which can be found by one dimensional search techniques (e.g., bisection method) \cite{WMMSE} or updated via sub-gradient method \cite{Subgrad}.

\subsection{PS Matrix Optimization with Fixed Transmit Beamforming}
\begin{figure*}
\begin{sequation}
\tag{20}
\begin{aligned}
    g(&\IRS)
= \sum\nolimits_{k=1}^K \s \big\{
\tr\big( \IRS^{\rH} \hHrk \mbf{C}_k \mbf{T}_k \mbf{C}_k^{\rH}\hHrk^{\rH}\IRS \hG\tW \hG^{\rH} \big)
+\eg \tr(\tW) \cdot \tr\big[ \IRS^{\rH} \hHrk \mbf{C}_k  \mbf{T}_k \mbf{C}_k^{\rH}\hHrk^{\rH}\IRS  \big]
\s +\erk\tr(\mbf{T}_k \mbf{C}_k^{\rH}\mbf{C}_k ) \cdot \tr\big(\IRS^{\rH}\IRS  \hG\tW \hG^{\rH} \big)
+\eg\erk\tr(\tW)  \\
&\hspace{0.15cm}\s\cdot\tr(\mbf{T}_k \mbf{C}_k^{\rH}\mbf{C}_k )\tr\big(\IRS^{\rH}\IRS\big)
\s + \tr\big(\hHrk \mbf{C}_k \mbf{T}_k \mbf{C}_k^{\rH} \hHdk^{\rH}\tW \hG^{\rH} \IRS^{\rH}  \big)
\s
+\tr\big(\hG \tW \hHdk \mbf{C}_k \mbf{T}_k \mbf{C}_k^{\rH} \hHrk^{\rH}\IRS\big)
-\tr\big(\hG \mbf{W}_{k}\mbf{T}_k\mbf{C}_k^{\rH} \hHrk^{\rH}\IRS\big)
-\tr\big(\hG^{\rH} \mbf{C}_{k}\mbf{T}_k\mbf{W}_k^{\rH} \hHrk\IRS^{\rH} \big) \big\}  \\
&\hspace{0.42cm}=\hspace{0.1cm}\s\tr \big( \IRS^{\rH}\mbf{A}_0\IRS\mbf{B}_0 \big) + \tr \big( \IRS^{\rH}\mbf{A}_1\IRS \big)+\tr \big( \IRS^{\rH}\IRS\mbf{B}_1 \big)
     +a_2 \cdot\tr \big( \IRS^{\rH} \IRS \big)+ \tr \big(\mbf{D}\IRS^{\rH} \big) + \tr \big( \mbf{D}^{\rH} \IRS\big)
\label{eq:func_g}
\end{aligned}
\end{sequation}
\hrulefill
\end{figure*}
Subsequently, we consider the second sub-problem of optimizing the PS matrix $\bm{\Theta}$ with the obtained precoder $\mlW^{*}$ and $\{\mbf{C}_k^{*}\},\{\mbf{T}_k^{*}\}$. For brevity, we omit the superscription of $\mlW^{*}$, $\{\mbf{C}_k^{*}\}$ and $\{\mbf{T}_k^{*}\}$ in this section. Denote $\bm{\phi}=[\phi_1,\cdots,\phi_N]^{\rT}$, we reconstruct $\mcal{P}(\text{B2})$ as follows.
\begin{Lem1}\label{prop:ABD}
$\mcal{P}(\text{B2})$ is equivalent to
\begin{sequation}
\setcounter{equation}{17}
\mcal{P}(\text{B2.1})\quad \underset{\scriptscriptstyle \bm{\phi}}{\text{min}} \hspace {0.2cm}  h(\bm{\phi}) \quad\text{s.t.} \hspace{0.3cm} {\phi_n \in \mcal{F}}, \hspace{0.15cm} \forall n \in \mcal{N} ,
\label{pro:B2_1}
\end{sequation}
where $ h(\bm{\phi})=(e^{j\bm{\phi}})^{\rH}\mbf{F}e^{j\bm{\phi}}+2\text{Re}\big\{ (e^{j\bm{\phi}})^{\rH} \mbf{d} \big\}$, and
\begin{sequation}
\begin{gathered}
     \mbf{F}=\mbf{F}_0+\mbf{F}_1+\mbf{F}_2+\mbf{F}_3,\hspace{0.15cm}
     \mbf{d}=\big[ [\mbf{D}]_{1,1},\cdots, [\mbf{D}]_{N,N} \big]^{\rT}, \\
     \mbf{F}_0=\mbf{A}_0\circ\mbf{B}_0^{\rT}, \hspace{0.15cm} \mbf{F}_1=\mbf{A}_1\circ\mbf{I}, \hspace{0.15cm} \mbf{F}_2=\mbf{I}\circ\mbf{B}_1^{\rT}, \hspace{0.15cm} \mbf{F}_3=a_2 \cdot \mbf{I},
\end{gathered}
\label{eq:Uz}
\end{sequation}
\begin{sequation}
\begin{aligned}
     &\mbf{A}_0= \sum\nolimits_{k=1}^K  \hat{\mbf{H}}_{r,k}\nCopt\nTopt\nCopt^{\rH}\hat{\mbf{H}}_{r,k}^{\rH}, \\
     &\mbf{A}_1= \sum\nolimits_{k=1}^K  \eg \cdot\tr(\ntWopt)\cdot \hHrk\nCopt\nTopt\nCopt^{\rH}\hHrk^{\rH}, \\
     &a_2 =\eg\erk\cdot\tr(\nTopt \nCopt^{\rH}\nCopt ) \cdot\tr(\ntWopt), \hspace{0.2cm}
     \mbf{B}_0=  \hG\ntWopt\hG^{\rH}, \\
     &\mbf{B}_1=  \sum\nolimits_{k=1}^K \erk \cdot\tr(\nTopt \nCopt^{\rH} \nCopt )\cdot \hG\ntWopt\hG^{\rH}, \\
     &\mbf{D}= \sum\nolimits_{k=1}^K   \big[ \hHrk\nCopt\nTopt \big( \nCopt^{\rH}\hHdk^{\rH}  \ntWopt \hG^{\rH} - \mbf{W}_{k}^{\rH}\hG^{\rH} \big)  \big].
\end{aligned}
\label{eq:ABD}
\end{sequation}
\end{Lem1}

\emph{Proof:} Firstly, we expand $\mbf{E_k}$ in $f(\IRS)$ of $\mcal{P}(\text{B2})$ by (\ref{eq:MSE}). Note that, in this sub-problem, $\IRS$ is the only variable, i.e., $\{\nTopt\}$, $\{\nCopt\}$ and $\nWopt$ are assumed to be constant. Thus, the terms $\sum_{k=1}^K \left\{ \tr[\nTopt(\sigma_{k}^2 \nCopt^{\rH}\nCopt +\mbf{I})]-\omega_k \log_2 | 1 / \omega_k \cdot\nTopt|\right\}$  are irrelevant to $\IRS$ and can be leaved out in $f(\IRS)$. Substituting $\hHk$ with $\hHk=\hHdk^{\rH}+\hHrk^{\rH}\IRS\hG$ and applying the lemma in \cite{ImCSI}, $f(\IRS)$ in $\mcal{P}(\text{B2})$ can be transformed into (\ref{eq:func_g}) at the top of this page, where $\mbf{A}_0,\mbf{A}_1,a_2,\mbf{B}_0,\mbf{B}_1, \mbf{D}$ are given by (\ref{eq:ABD}).

Utilizing the matrix identity in \cite{mat}, we have
\begin{sequation}
\begin{aligned}
    \tr \big( \IRS^{\rH}\mbf{A}_0\IRS\mbf{B}_0 \big) = (e^{j\bm{\phi}})^{\rH}(\mbf{A}_0\circ\mbf{B}_0^{\rT})e^{j\bm{\phi}},\\
    \tr \big(\mbf{D}\IRS^{\rH} \big) = \mbf{d}^{\rH}e^{j\bm{\phi}} , \hspace{0.2cm}
    \tr \big( \mbf{D}^{\rH} \IRS\big) = (e^{j\bm{\phi}})^{\rH} \mbf{d}.
\end{aligned}
\label{eq:h_Uz}
\end{sequation}
Transforming the other terms in (\ref{eq:func_g}) similarly, $g(\IRS)$ can be rewritten as \cite{JP}
\begin{sequation}
\begin{aligned}
h(\bm{\phi})&=
(e^{j\bm{\phi}})^{\rH}(\mbf{F}_0+\mbf{F}_1+\mbf{F}_2+\mbf{F}_3)e^{j\bm{\phi}}+2\text{Re}\big\{ (e^{j\bm{\phi}})^{\rH} \mbf{d} \big\} \\
&=(e^{j\bm{\phi}})^{\rH}\mbf{F}e^{j\bm{\phi}}+2\text{Re}\big\{ (e^{j\bm{\phi}})^{\rH} \mbf{d} \big\},
\end{aligned}
\end{sequation}
where $\mbf{F}_0,  \mbf{F}_1,  \mbf{F}_2, \mbf{F}_3, \mbf{F}$ and $\mbf{d}$ are given by (\ref{eq:Uz}),
which completes the proof.\hfill$\square$

\subsubsection{MM-Based Solution}
In this part, we resort to the MM method for solving $\mcal{P}(\text{B2.1})$. According to \cite{MM}, the surrogate function $\hat{h}(\bm{\phi}|\bm{\phi}^r)$ for the quadratic function $h(\bm{\phi})$ at the $r^{th}$ iteration can be constructed as
\begin{sequation}
\begin{aligned}
     \hat{h}(\bm{\phi}|\bm{\phi}^r) =& \mu (e^{j\bm{\phi}})^{\rH}e^{j\bm{\phi}}
     +(e^{j\bm{\phi}^r})^{\rH}\left(\mu\mbf{I}-\mbf{F}\right)e^{j\bm{\phi}^r} \\
     -2&\text{Re}\big\{(e^{j\bm{\phi}})^{\rH} \left(\mu\mbf{I}-\mbf{F}\right)e^{j\bm{\phi}^r} \big\}
     +2\text{Re}\big\{ (e^{j\bm{\phi}})^{\rH} \mbf{d} \big\},
\end{aligned}
\end{sequation}
where $\mu$ is the maximum eigenvalue of matrix $\mbf{F}$. Obviously, since $(e^{j\bm{\phi}})^{\rH}e^{j\bm{\phi}}=N$, minimizing $\hat{h}(\bm{\phi}|\bm{\phi}^r)$ is equivalent to maximizing $\text{Re}\big\{ (e^{j\bm{\phi}})^{\rH}\mbf{z}^r \big\}$, where
\begin{sequation}
     \mbf{z}^r=\left(\mu\mbf{I}-\mbf{F}\right)e^{j\bm{\phi}^r}-\mbf{d}.
\label{eq:q}
\end{sequation}
 Therefore, the optimal solution for $\bm{\phi}^{r+1}$ can be obtained by
\begin{sequation}
     \bm{\phi}^{r+1}=e^{j\angle \mbf{z}^r}.
\label{eq:theta}
\end{sequation}
Due to the phase shift constraint of $\phi_n \in \mcal{F}$, we need an additional quantization operation, where $\phi_n$ is updated as
\begin{sequation}
\hat{\phi}_n = \underset{\psi \in \mcal{F}}{\arg \min} \left| \phi_n - \psi \right| ,\hspace{0.2cm} \forall n\in \mcal{N}.
\label{eq:quantize}
\end{sequation}

\subsubsection{SCA-Based Solution}
In this part, we make use of the SCA technique to tackle the surrogate function of $\mcal{P}(\text{B2.1})$ instead. By applying the second order Taylor expansion \cite{Taylor}, the approximation of $h(\bm{\phi})$ around $\bm{\phi}^{r}$ can be expressed as
\begin{sequation}
     \hat{h}(\bm{\phi}|\bm{\phi}^{r})=h(\bm{\phi}^{r})+\nabla h(\bm{\phi}^{r})^{\rT}(\bm{\phi}-\bm{\phi}^{r}) +\frac{\beta_r}{2}\|\bm{\phi}-\bm{\phi}^{r} \|^2,
\end{sequation}
where $\beta_r$ should be chosen to satisfy $\hat{h}(\bm{\phi}|\bm{\phi}^{r})\geq h(\bm{\phi})$ \cite{beta}, which can be determined by Armijo rule \cite{Arm}, and the gradient is given by
\begin{sequation}
     \nabla h(\bm{\phi}^{r})=2\text{Re} \big\{ -j e^{-j\bm{\phi}^{r}} \circ (\mbf{F} e^{j\bm{\phi}^{r}}+\mbf{d}) \big\}.
\label{eq:grad}
\end{sequation}
In order to minimize $\hat{h}(\bm{\phi}|\bm{\phi}^{r})$, we apply the gradient descent method. Thus, $\bm{\phi}$ is updated by
\begin{sequation}
     \bm{\phi}^{r+1}=\bm{\phi}^{r}- \nabla h(\bm{\phi}^{r}) / \beta_r.
\label{eq:phi}
\end{sequation}

To sum up, the overall algorithm with the proposed two passive solutions is concluded in Algorithm \ref{alg:Main}. The convergence analysis of the algorithm is omitted here due to space limitation, which can be can be found in \cite{WSR_hyg}, \cite{BCD} and \cite{JP}.
% Add ref

\begin{algorithm}
\caption{Joint Beamforming Design Method}\label{alg:Main}
\begin{algorithmic}[1]
\STATE \tbf{Initialize:}  $\IRS^0$, $\mlW^0$. Set iteration index $r=0$.
\REPEAT
\STATE Given $\IRS^r$ and $\mlW^r$, update $\{\mbf{C}_k^{r}\}$, $\{\mbf{E}_k^{r}\}$ and $\{\mbf{T}_k^{r}\}$ by (\ref{eq:com}) and (\ref{eq:Ek_Tk}), respectively.
\STATE Given $\IRS^r$, determine $\lambda$ and update $\{\mW_{k}^{r+1}(\lambda)\}$ by (\ref{eq:Wk}) with the obtained $\{\mbf{C}_k^{r}\}$ and $\{\mbf{T}_k^{r}\}$.
\STATE Compute $\mbf{d}$ and $\mbf{F}$ by (\ref{eq:Uz}) with the obtained $\{\mbf{C}_k^{r}\}$, $\{\mbf{T}_k^{r}\}$ and $\mlW^{r+1}$.
\STATE Update $\bm{\phi}^{r+1}$ by (\ref{eq:theta}) or (\ref{eq:phi}).
\STATE Quantize the elements of $\bm{\phi}^{r+1}$ by (\ref{eq:quantize}).
\STATE Construct $\IRS^{r+1}$ with the obtained $\bm{\phi}^{r+1}$.
\STATE Set $r=r+1$.
\UNTIL {the fractional decrease of $\mcal{P}(\text{B2})$ is below a predefined threshold.}
\STATE \tbf{Output:} $\IRS^{r+1},\mlW^{r+1}$.
\end{algorithmic}
\end{algorithm}

\section{Numerical Results}
\begin{figure}
    \centering
    \includegraphics[width=0.35\textwidth]{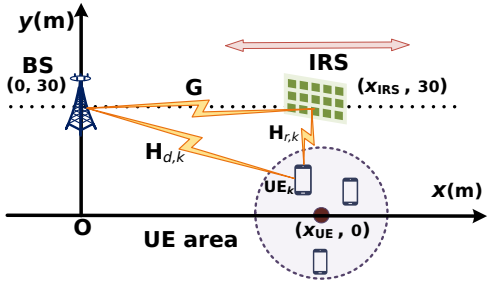}
    \caption{The simulated IRS-aided MU MIMO communication system.}
    \label{fig:simu}
\end{figure}

In this section, simulation results are provided to validate the effectiveness of the proposed schemes. As shown in Fig. \ref{fig:simu}, we consider the IRS-aided MU-MIMO communication system consisted of one BS equipped with 8 transmit antennas, one IRS with 100 reflecting elements, and 3 multi-antenna users, each with 2 receive antennas. We set the number of the data streams for each user as $N_s=2$. The transmission bandwidth and noise power spectral density are set as 180 kHz and $-170$ dBm/Hz, respectively. We assume that the BS and IRS are located at (0, 30) m and ($x_{IRS}$, 30) m, respectively. The users are randomly distributed in a circle centered at ($x_{UE}$, 0) m with radius 10 m. Firstly, we set $x_{IRS}=x_{UE}=200$ m, if not specified otherwise. The weights $\{\omega_k\}$ for the rate are set to be proportional to the inverse of the corresponding direct-link path loss. The transmit power constraint $P_t$ is set as 0 dBm if not specified otherwise. All the simulation results are obtained by averaging over 1000 independent channel realizations.

For the estimated channel, we adopt the Rayleigh fading model for the direct link $\hHdk$ whose path loss in dB is $32.6+36.7 \lg d$ \cite{3GPP}, and Rician fading model for the BS-IRS link $\hG$ and IRS-users $k$ link $\hHrk$, which is given by
\begin{sequation}
     \hG=\kappa_G \left(\sqrt{\frac{\nu}{\nu+1}}\hG^{\text{LOS}}+\sqrt{\frac{1}{\nu+1}}\hG^{\text{NLOS}} \right),
\end{sequation}
where $\kappa_G=35.6+22.0 \lg d$ (dB) is the corresponding path loss, $\nu=10$ is the Rician factor. $\hG^{\text{LOS}}$ represents the light of sight (LOS) components of the IRS-aided channels. We assume that the antennas at the BS and the passive reflecting elements at the IRS are both arranged in a half-wavelength uniform linear array (ULA). Thus, we have $\hG^{\text{LOS}}=\mbf{a}_r(\varphi_r)\mbf{a}_t^{\rH}(\varphi_t)$, where $\mbf{a}$ is the antenna steering vector, $\varphi_r$ and $\varphi_t$ are the angular parameters. $\hG^{\text{NLOS}}$ stands for the non-LOS (NLOS) component, which follows Rayleigh fading. $\hHrk$ is defined in the same manner.

We assume that the channel estimation errors follow the i.i.d. zero mean CSCG distribution and share the same normalized MSE (NMSE), which is defined as
\begin{sequation}
     \varrho= \mbb{E}\big[|\mbf{H}-\widehat{\mbf{H}}|^2 \big]/\mbb{E}\big[|\widehat{\mbf{H}}|^2 \big].
\end{sequation}

The annotations for the ensuing curves are as follows:
\begin{enumerate}
  \item \textbf{MM:} Optimizations with the MM-based passive solution, i.e., update $\bm{\phi}^{r+1}$ by (\ref{eq:theta}) in Algorithm \ref{alg:Main}.
  \item \textbf{SCA:} Optimizations with the SCA-based passive solution, i.e., update $\bm{\phi}^{r+1}$ by (\ref{eq:phi}) in Algorithm \ref{alg:Main}.
  \item \textbf{Fixed IRS:} Given the initialized random $\IRS^0$, optimizations with the active beamforming only.
  \item \textbf{No IRS:} Direct transmissions only.
\end{enumerate}

\begin{figure} \centering
\includegraphics[width=0.7\columnwidth]{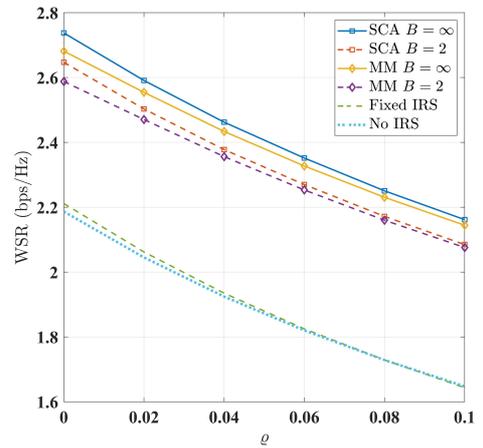}
\caption{WSR versus $\varrho$. ($P_t=0$ dBm)}
\label{fig:rho}
\end{figure}
Fig. \ref{fig:rho} presents the WSR of the system with different channel estimation errors. Obviously, the performance of all the schemes declines when the available CSI condition is getting worse. From the figure, we can also find that, if the PSs of the IRS are not optimized, the performance gain brought by the IRS is very small. Meanwhile, we observe a significant improvement of the WSR with the MM and SCA based scheme, verifying the effectiveness of the proposed designs. Note that, we also take the impact of discrete PSs at the IRS into account. It can be seen that, the performance of the proposed joint design schemes with 2-bit quantization for the PSs of the IRS is close to the continuous PSs case. Hence, we emphasize the performance of applying the discrete PSs for the IRS in the sequel. Furthermore, it is observed that, the SCA-based scheme outperforms than MM-based scheme under different conditions, indicating its superiority in consideration.

\begin{figure} \centering
\includegraphics[width=0.7\columnwidth]{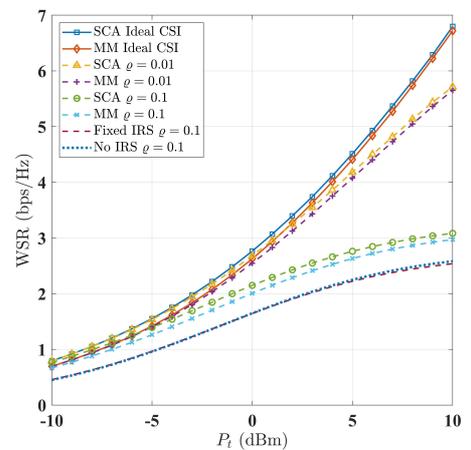}
\caption{WSR versus $P_t$. ($B=2$)}
\label{fig:Pt}
\end{figure}
In Fig. \ref{fig:Pt}, we investigate the WSR as a function of transmit power $P_t$. Obviously, all the schemes perform better with larger transmit power. Still, we can find that both the SCA and MM based designs achieve impressive performance gain in all cases compared with the fixed IRS scheme, verifying the viability of our proposed two schemes. However, it is worth noting that, the performance gap between different CSI conditions is widening with the increase of transmit power $P_t$, which calls for more careful design in the future study.

\begin{figure} \centering
\includegraphics[width=0.7\columnwidth]{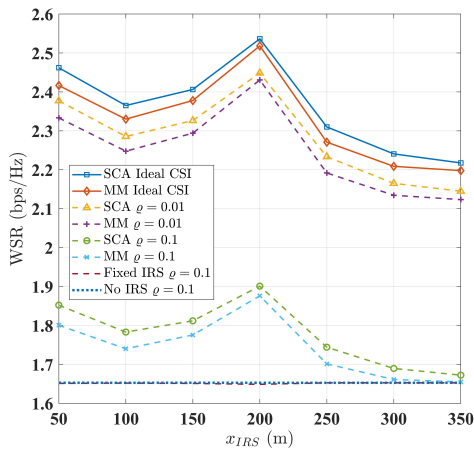}
\caption{WSR versus $x_{IRS}$. ($B=2,x_{UE}=200$ m)}
\label{fig:xIRS}
\end{figure}
In Fig. \ref{fig:xIRS}, the performance with the different horizontal position of the IRS is plotted. Apparently, we can see that, when the horizontal coordinate of the IRS moves from 50 m to 350 m, the WSR first falls down, then goes up and reaches its maximal value when IRS is deployed at (200, 30) m. After that, it begins declining again. This is mainly resulted from the joint impact of the path losses of the channels $\mbf{H}_{r,k}$ and $\mbf{G}$, which casts interesting insight on the deployment of IRS.

\section{Conclusions}
In this paper, we have investigated the joint beamforming design in IRS assisted MU-MIMO downlink transmissions with multi-antenna users and channel estimation errors. We have formulated the optimization problem aiming at maximizing the WSR and transformed it into an equivalent WMMSE minimization problem. To deal with the coupled variables in the optimization problem, we have utilized the BCD method. We have decomposed the original joint design problem into two sub-problems and used the Lagrange multipliers method for the first sub-problem. For the second sub-problem, we have proposed two solutions, namely MM-based algorithm and SCA-based technique. To fit the constraints of low-resolution PSs, we have quantized the optimal infinite angles into a discrete set in the iterative optimization process directly. Through simulation results, we have demonstrated the effectiveness of the proposed two schemes, especially the superiority of the SCA-based method in various settings of IRS-aided MU-MIMO communication systems with multi-antenna users and channel estimation errors.

%\end{multicols}

\end{document}